\documentclass[aps,prd,amsfonts,amssymb,amsmath,twocolumn,superscriptaddress,showpacs,eqsecnum,floatfix]{revtex4}

\usepackage{graphicx}
\usepackage{bm}

\newcommand{\bmath}[1]{\mbox{{\boldmath{{$#1$}}}}}

\begin{document}

\title{Soliton and black hole solutions of ${\mathfrak {su}}(N)$ Einstein-Yang-Mills
theory in anti-de Sitter space}

\author{J. E. Baxter}
\affiliation{Department of Applied Mathematics, The University of
Sheffield, Hicks Building, Hounsfield Road, Sheffield, S3 7RH,
United Kingdom.}
\author{Marc Helbling}
\affiliation{INSA de Rouen, Laboratoire de Math\'ematiques (LMI),
Place Emile Blondel BP 08, 76131 Mont Saint Aignan Cedex, France.}
\author{Elizabeth Winstanley} \email{E.Winstanley@sheffield.ac.uk}
\affiliation{Department of Applied Mathematics, The University of
Sheffield, Hicks Building, Hounsfield Road, Sheffield, S3 7RH,
United Kingdom.}

\date{\today }

\begin{abstract}
We present new soliton and hairy black hole solutions of ${\mathfrak {su}}(N)$ Einstein-Yang-Mills theory
in asymptotically anti-de Sitter space.
These solutions are described by $N+1$ independent parameters, and have $N-1$
gauge field degrees of freedom.
We examine the space of solutions in detail for ${\mathfrak {su}}(3)$ and ${\mathfrak {su}}(4)$ solitons
and black holes.
If the magnitude of the cosmological constant is sufficiently large, we find solutions where
all the gauge field functions have no zeros.
These solutions are of particular interest because we anticipate that at least some of them will
be linearly stable.
\end{abstract}


\maketitle

\section{Introduction}
\label{sec:intro}

Soliton and hairy black hole solutions of Einstein-Yang-Mills (EYM) theory
and its variants have been the subject of
extensive research since the discovery of non-trivial, spherically symmetric solitons
\cite{Bartnik} and `colored' black holes \cite{Bizon} in ${\mathfrak {su}}(2)$ EYM in asymptotically flat space-time.
These black holes are `hairy' in the sense that they have no magnetic charge, and are therefore indistinguishable
at infinity from a standard Schwarzschild black hole.
There are discrete families of solutions, indexed by the event horizon radius $r_{h}$ (with $r_{h}=0$ for
solitons) and $n$, the number of zeros of the single gauge field function $\omega $, each pair
$\left( r_{h},n \right)$ identifying a solution of the field equations.
A key feature of the solutions is that $n>0$, so that the gauge field function must have at least one zero
(or `node').
These solutions, while they violate the `letter' of the no-hair conjecture, may be thought of
as not contradicting its `spirit', since they are found to be unstable to classical, linear, spherically
symmetric perturbations \cite{Straumann}.
There is also a large literature concerning analytic studies of the
asymptotically flat ${\mathfrak {su}}(2)$ EYM field equations
\cite{BFM,Smoller},
proving the existence of the above numerical solutions and detailed properties of the phase space of solutions.
Since these initial discoveries
a plethora of new soliton and black hole solutions have been found (see \cite{Volkov} for a review).
The present work combines two natural extensions of these initial studies: the generalization to
${\mathfrak {su}}(N)$ EYM, and the inclusion of a negative cosmological constant.
We now briefly review each of these generalizations in turn.

Firstly, in asymptotically flat space, both charged and neutral numerical solutions
of the ${\mathfrak {su}}(N)$ EYM field equations have been found
\cite{Galtsov}.
We consider in this paper only
purely magnetic solutions, which, in the asymptotically flat case,
are described by $N-1$ gauge field functions $\omega _{j}$.
As in asymptotically flat ${\mathfrak {su}}(2)$ EYM,
solutions exist at discrete points in the parameter space, and
can be indexed by the radius of the event horizon (if there is one) and the number of nodes of the $\omega _{j}$
(all $\omega _{j}$ having at least one zero).
Once again, there is a general result \cite{Brodbeck} that all these solutions
must be unstable.
The ${\mathfrak {su}}(N)$ EYM field equations are considerably more complicated than those for
${\mathfrak {su}}(2)$ and correspondingly less analytic work has been done.
Local existence of solutions of the field equations near the origin (for solitons), black hole event
horizon (if there is one) and at infinity has been established \cite{Kunzle1,Oliynyk}.
There is a heuristic proof (following \cite{BFM}) of the existence of black hole solutions for
general $N$ \cite{ewsuN}, but more rigorous work exits only for the case of ${\mathfrak {su}}(3)$
\cite{Ruan}.

The second generalization of asymptotically flat ${\mathfrak {su}}(2)$ EYM that we consider in this
paper is the inclusion of a non-zero cosmological constant $\Lambda $.
When the cosmological constant is positive, soliton \cite{su2poslambda} and black hole \cite{Torii}
${\mathfrak {su}}(2)$ solutions have been found.
These solutions possess a cosmological horizon and approach de Sitter space at infinity
(for a complete classification of the possible space-time structures, see \cite{BFM1}).
The phase space of solutions is again discrete, and the single gauge field function
$\omega $ must have at least one zero.
Unsurprisingly, these solutions again turn out to be unstable \cite{Torii,Brodbeck3}.
The inclusion of a negative cosmological constant (so that the space-time is asymptotically anti-de Sitter (adS))
may be motivated by recent progress in string theory,
particularly the adS/CFT correspondence \cite{Maldacena}.
It is found \cite{ew1,Bjoraker} that the solutions of ${\mathfrak {su}}(2)$ EYM in adS possess
quite different properties compared with their asymptotically flat or asymptotically de Sitter cousins.
In particular, solutions for which the
gauge field function $\omega $ has no zeros exist for sufficiently large $\left| \Lambda \right| $.
Solutions exist in continuous open subsets of the parameter space, rather than at discrete points.
In addition, for sufficiently large $\left| \Lambda \right| $, at least some of these
solutions are stable under linear, spherically symmetric perturbations
\cite{ew1,Bjoraker} (this was subsequently extended
to cover non-spherically symmetric linear perturbations in \cite{Sarbach}).
Therefore, while black holes cannot be given stable YM hair in either asymptotically flat or asymptotically
de Sitter space,
in asymptotically anti-de Sitter space, stable gauge field hair is possible.

We are thus led to the following natural question: are there stable solutions of the ${\mathfrak {su}}(N)$
EYM solutions with a negative cosmological constant?
In this paper we will present the first soliton and hairy black hole solutions of ${\mathfrak {su}}(N)$ EYM
in adS, for $N>2$.
We consider only purely magnetic gauge fields, so that the YM field is described by $N-1$ functions
$\omega _{j}$.
We make a detailed study of the phase space of solutions and their general properties in the particular
cases $N=3$, $4$.
Of particular interest is the existence, for sufficiently large $\left| \Lambda \right| $,
of solutions in which all the $\omega _{j}$ have no zeros.
We anticipate that at least some of these solutions will be stable under linear, spherically symmetric perturbations,
and will examine their stability in detail elsewhere.

The outline of this paper is as follows.
In section \ref{sec:equations} we discuss the field equations, our ansatze for the fields and the boundary
conditions that must be satisfied, considering the cases of black holes and solitons separately.
Our new solutions are discussed in detail in section \ref{sec:solutions},
focussing particularly on the phase space of solutions.
Finally, our conclusions can be found in section \ref{sec:conc}.
Throughout this paper, the metric has signature $(-,+,+,+)$ and we use units in which $4\pi G=c=1$.

\section{Ansatz, field equations and boundary conditions}
\label{sec:equations}

\subsection{Ansatz and field equations}
\label{sec:ansatz}

We consider static, spherically symmetric, four-dimensional solitons and black holes with metric
\begin{equation}
ds^{2} = - \mu S^{2} \, dt^{2} + \mu ^{-1} \, dr^{2} +
r^{2} \, d\theta ^{2} + r^{2} \sin ^{2} \theta \, d\phi ^{2} ,
\label{eq:metric}
\end{equation}
where the metric functions $\mu $ and $S$ depend on the radial co-ordinate $r$ only.
In the presence of a negative cosmological constant $\Lambda <0$, we write the metric function $\mu $ as
\begin{equation}
\mu (r) = 1 - \frac {2m(r)}{r} - \frac {\Lambda r^{2}}{3}.
\end{equation}
The most general, spherically symmetric, ansatz for the ${\mathfrak {su}}(N)$ gauge potential is \cite{Kunzle}:
\begin{eqnarray}
A & = &
{\cal {A}} \, dt + {\cal {B}} \, dr +
\frac {1}{2} \left( C - C^{H} \right) \, d\theta
\nonumber \\ & &
- \frac {i}{2} \left[ \left(
C + C^{H} \right) \sin \theta + D \cos \theta \right] \, d\phi ,
\label{eq:gaugepot}
\end{eqnarray}
where ${\cal {A}}$, ${\cal {B}}$, $C$ and $D$ are all $\left( N \times N \right) $ matrices and
$C^{H}$ is the Hermitian conjugate of $C$.
The matrices ${\cal {A}}$ and ${\cal {B}}$ are purely imaginary, diagonal, traceless and depend only
on the radial co-ordinate $r$.
The matrix $C$ is upper-triangular, with non-zero entries only immediately above the diagonal:
\begin{equation}
C_{j,j+1}=\omega_j (r) e^{i\gamma _{j}(r)},
\end{equation}
for $j=1,\ldots,N-1$.
In addition, $D$ is a constant matrix:
\begin{equation}
D=\mbox{Diag}\left(N-1,N-3,\ldots,-N+3,-N+1\right) .
\label{eq:matrixD}
\end{equation}
Here we consider only purely magnetic solutions, so we set ${\cal {A}} \equiv 0$.
We may also take ${\cal {B}}\equiv 0$ by a choice of gauge \cite{Kunzle}.
From now on we will assume that all the $\omega _{j}(r)$ are non-zero
(see, for example, \cite{Galtsov} for the possibilities in asymptotically flat space if this
assumption does not hold).
In this case one of the Yang-Mills equations becomes \cite{Kunzle}
\begin{equation}
\gamma _{j} = 0 \qquad \forall j=1,\ldots , N-1.
\end{equation}
Our ansatz for the Yang-Mills potential therefore reduces to
\begin{equation}
A = \frac {1}{2} \left( C - C^{H} \right) \, d\theta - \frac {i}{2} \left[ \left(
C + C^{H} \right) \sin \theta + D \cos \theta \right] \, d\phi ,
\label{eq:gaugepotsimple}
\end{equation}
where the only non-zero entries of the matrix $C$ are
\begin{equation}
 C _{j,j+1} = \omega _{j}(r).
\end{equation}
The gauge field is therefore described by the $N-1$ functions $\omega _{j}(r)$.
We comment that our ansatz (\ref{eq:gaugepotsimple}) is by no means the only possible choice in
${\mathfrak {su}}(N)$ EYM.
Techniques for finding {\em {all}} spherically symmetric ${\mathfrak {su}}(N)$ gauge potentials
can be found in \cite{Bartnik1}, where all irreducible models are explicitly listed for $N\le 6$.

With the ansatz (\ref{eq:gaugepotsimple}), there are $N-1$ non-trivial Yang-Mills equations
for the $N-1$ functions $\omega _{j}$:
\begin{equation}
r^2\mu\omega''_{j}+\left(2m-2r^3 p_{\theta}-\frac{2\Lambda
r^3}{3}\right)\omega'_{j}+W_j\omega_j=0
\label{eq:YMe}
\end{equation}
for $j=1,\ldots,N-1$, where a prime $'$ denotes $d/dr $,
\begin{eqnarray}
p_{\theta}&=&
\frac{1}{4r^4}\sum^N_{j=1}\left[\left(\omega^2_j-\omega^2_{j-1}-N-1+2j\right)^2\right],
\label{eq:ptheta}
\\
W_j&=&
1-\omega^2_j+\frac{1}{2}\left(\omega^2_{j-1}+\omega^2_{j+1}\right),
\end{eqnarray}
and $\omega_0=\omega_N=0$.
The Einstein equations take the form
\begin{equation}
m' =
\mu G+r^2p_{\theta},
\qquad
\frac{S'}{S}=\frac{2G}{r},
\label{eq:Ee}
\end{equation}
where
\begin{equation}
G=\sum^{N-1}_{j=1}\omega_j'^2.
\label{eq:Gdef}
\end{equation}
Altogether, then, we have $N+1$ ordinary differential equations for the $N+1$ unknown functions $m(r)$, $S(r)$
and $\omega _{j}(r)$.

The field equations (\ref{eq:YMe},\ref{eq:Ee}) are invariant under the transformation
\begin{equation}
\omega _{j} (r) \rightarrow -\omega _{j} (r)
\label{eq:omegaswap}
\end{equation}
for each $j$ independently, and also under the substitution:
\begin{equation}
j \rightarrow N - j.
\label{eq:Nswap}
\end{equation}

\subsection{Boundary conditions}
\label{sec:bcs}

The field equations (\ref{eq:YMe},\ref{eq:Ee}) are singular at the origin $r=0$ (for regular, soliton solutions),
the black hole event horizon $r=r_{h}$ (if there is one) and at infinity $r\rightarrow \infty $.
We therefore now discuss the boundary conditions that must be satisfied by the field variables at these singular points.
Local existence of solutions of the field equations in neighborhoods of these singular points will be rigorously proved
elsewhere \cite{BW}, generalizing the local existence proofs in the asymptotically flat case \cite{Kunzle1,Oliynyk}.
The boundary conditions satisfied by black hole solutions are more easily stated, so we consider those first.

\subsubsection{Black holes}
\label{sec:bcbhs}

We assume there is a regular, non-extremal, black hole event horizon at $r=r_{h}$, where $\mu (r)$ has
a single zero.
This fixes the value of $m(r_{h})$ to be:
\begin{equation}
2m( r_{h} ) = r_{h} - \frac {\Lambda r_{h}^{3}}{3}.
\end{equation}
The field variables $\omega _{j}(r)$, $m(r)$ and $S(r)$ will have regular Taylor series expansions about $r=r_{h}$:
\begin{eqnarray}
m(r) & = & m (r_{h}) + m' (r_{h}) \left( r - r_{h} \right)
+ O \left( r- r_{h} \right) ^{2} ;
\nonumber \\
\omega _{j} (r) & = & \omega _{j}(r_{h}) + O \left( r -r_{h} \right) ;
\nonumber \\
S(r) & = & S(r_{h}) + O\left( r - r_{h} \right) .
\label{eq:horizon}
\end{eqnarray}
Setting $\mu (r_{h})=0$ in the Yang-Mills equations (\ref{eq:YMe}) fixes the derivatives of the
gauge field functions at the horizon:
\begin{equation}
\omega _{j} ' (r_{h}) = - \frac {W_{j}(r_{h})\omega _{j}(r_{h})}{2m(r_{h}) - 2r_{h}^{3} p_{\theta } (r_{h})
-\frac {2\Lambda r_{h}^{3}}{3}}.
\end{equation}
Therefore the expansions (\ref{eq:horizon}) are determined by
the $N+1$ quantities $\omega _{j}(r_{h})$, $r_{h}$, $S(r_{h})$ for fixed
cosmological constant $\Lambda $.
For the event horizon to be non-extremal, it must be the case that
\begin{equation}
2m'(r_{h}) = 2r_{h}^{2} p_{\theta} (r_{h}) < 1- \Lambda r_{h}^{2},
\label{eq:constraint}
\end{equation}
which weakly constrains the possible values of the gauge field functions $\omega _{j}(r_{h}) $
at the event horizon.
Since the field equations (\ref{eq:YMe},\ref{eq:Ee}) are invariant under the transformation
(\ref{eq:omegaswap}), we may consider
$\omega _{j}(r_{h}) >0$ without loss of generality.

At infinity, the boundary conditions are considerably less stringent than in the asymptotically flat case.
In order for the metric (\ref{eq:metric}) to be asymptotically adS,
we simply require that the field variables $\omega _{j}(r)$, $m(r)$ and $S(r)$ converge to constant values as
$r\rightarrow \infty $, and have regular Taylor series expansions in $r^{-1}$ near infinity:
\begin{eqnarray}
m(r)  & = &  M + O \left( r^{-1} \right) ;
\qquad
S(r) = 1 + O\left( r^{-1} \right) ;
\nonumber \\
\omega _{j}(r) & = & \omega _{j,\infty } + O \left( r^{-1} \right) .
\label{eq:infinity}
\end{eqnarray}
Since $\Lambda <0$, there is no cosmological horizon.

\subsubsection{Solitons}
\label{sec:bcsolitons}

Soliton solutions have the same boundary conditions (\ref{eq:infinity}) as $r\rightarrow \infty $ as
black hole solutions.
The boundary conditions at a regular origin, however,
are considerably more complicated than at a black hole event horizon or at infinity.
For the asymptotically flat case, they have been derived in \cite{Kunzle1}.
As may be expected, the modifications required by the presence of a non-zero cosmological constant
are not great.
However, given the complexity of these boundary conditions, we now describe their derivation in some detail.

We begin by assuming a regular Taylor series expansion for all field variables near $r=0$:
\begin{eqnarray}
m(r) & = & m_{0} + m_{1}r + m_{2}r^{2} + O(r^{3});
\nonumber \\
S(r) & = & S_{0} + S_{1}r +S_{2}r^{2} + O(r^{3});
\nonumber \\
\omega _{j}(r) & = & \omega _{j,0} + \omega _{j,1}r + \omega _{j,2} r^{2} + O(r^{3});
\label{eq:origin1}
\end{eqnarray}
where the $m_{i}$, $S_{i}$ and $\omega _{j,i}$ are constants.
The expansions (\ref{eq:origin1}) are substituted into the field equations (\ref{eq:YMe},\ref{eq:Ee}) to
determine the values of the constants.
The constant $S_{0}$ is non-zero in order for the metric to be regular at the origin,
but otherwise arbitrary since the field equations involve only derivatives of $S$.

Regularity of the metric and curvature at the origin immediately gives:
\begin{equation}
m_{0}=m_{1}=m_{2} =0, \qquad S_{1}=0, \qquad \omega _{j,1}=0
\end{equation}
and
\begin{equation}
\omega _{j,0} =  \pm {\sqrt {j(N-j)}}.
\label{eq:omegaorigin}
\end{equation}
Without loss of generality (due to (\ref{eq:omegaswap})),
we take the positive square root in (\ref{eq:omegaorigin}).

Examination of the leading order terms in the Yang-Mills equations (\ref{eq:YMe}) gives the following
constraint on ${\bmath {\omega }}_{2} = \left( \omega _{1,2}, \ldots , \omega _{N-1,2} \right) ^{T}$:
\begin{equation}
{\cal {M}}_{N-1} {\bmath {\omega }}_{2} = 2{\bmath {\omega }}_{2}.
\end{equation}
\begin{widetext}
Here, ${\cal {M}}_{N-1}$ is the $(N-1)\times (N-1)$ matrix
\begin{equation}
{\cal {M}}_{N-1} = \left(
\begin{array}{ccccc}
2(N-1)&-\sqrt{(N-1)2(N-2)}&0&\cdots& 0 \\
-\sqrt{(N-1)2(N-2)}&2.2(N-2)&-\sqrt{2(N-2)3(N-3)}&\cdots&0 \\
0&-\sqrt{2(N-2)3(N-3)}&2.3(N-3)&\cdots & 0\\
\vdots&\vdots&\vdots&\ddots & \vdots\\
0 & 0 & 0 & \cdots  & -\sqrt{(N-1)2(N-2)}\\
0 & 0 & 0 & \cdots  & 2(N-1)
\end{array}
\right)
\label{eq:calM}
\end{equation}
Therefore ${\bmath {\omega }}_{2}$ is an eigenvector of the matrix ${\cal {M}}_{N-1}$ with eigenvalue $2$ if one
exists, otherwise ${\bmath {\omega }}_{2}={\bmath {0}}$.

To find the eigenvalues and eigenvectors of ${\cal {M}}_{N-1}$, we first note that it can be written in the form
\begin{equation}
{\cal {M}}_{N-1} = {\cal {D}}_{N-1}{\tilde {\cal {M}}}_{N-1}{\cal {D}}_{N-1}^{-1},
\end{equation}
where
\begin{eqnarray}
{\cal {D}}_{N-1} & = &
\mbox{Diag} \left( {\sqrt {N-1}}, {\sqrt {2(N-2)}}, {\sqrt {3(N-3)}}, \ldots , {\sqrt {N-1}} \right)
\nonumber
\\
{\tilde {\cal {M}}}_{N-1} & = &
\left(
\begin{array}{ccccc}
2(N-1)&-2(N-2)&0&\cdots& 0 \\
-(N-1)&2.2(N-2)&-3(N-3)&\cdots&0 \\
0&-2(N-2)&2.3(N-3)&\cdots & 0\\
\vdots&\vdots&\vdots&\ddots & \vdots\\
0 & 0 & 0 & \cdots  & -(N-1)\\
0 & 0 & 0 & \cdots  & 2(N-1)
\end{array}
\right) .
\end{eqnarray}
\end{widetext}
Then the matrices ${\cal {M}}_{N-1}$ and ${\tilde {\cal {M}}}_{N-1}$ have the same eigenvalues, and the eigenvectors of
${\cal {M}}_{N-1}$ can be deduced from those of ${\tilde {\cal {M}}}_{N-1}$.
This result is useful because the matrix ${\tilde {\cal {M}}}_{N-1}$ has been studied in detail in \cite{Kunzle1}.
There it is proved that the eigenvalues of ${\tilde {\cal {M}}}_{N-1}$ (and therefore those of ${\cal {M}}_{N-1}$)
are:
\begin{equation}
{\cal {E}}_{k} = k(k+1), \qquad k=1,\ldots, N-1 .
\label{eq:eigenvalues}
\end{equation}
The eigenvectors of ${\tilde {\cal {M}}}_{N-1}$ in general involve Hahn polynomials \cite{Hahn}, and can
be found explicitly in \cite{Kunzle1}.
The eigenvectors ${\bmath {v}}_{k}$
for $N=3,4$ will be presented in sections \ref{sec:su3solitons} and \ref{sec:su4solitons}
when we discuss the soliton solutions for ${\mathfrak {su}}(3)$ and ${\mathfrak {su}}(4)$ EYM, respectively.

Therefore, we set
\begin{equation}
{\bmath {\omega }}_{2}=b_{1}{\bmath {v}}_{1},
\end{equation}
where ${\bmath {v}}_{1}$ is a (suitably normalized)
eigenvector of ${\cal {M}}_{N-1}$ with eigenvalue $2$, and $b_{1}$ is an arbitrary constant.
From the Einstein equations (\ref{eq:Ee}), we find that $m_{3}$ and $S_{2}$ are fixed and given in terms of the
$\omega _{j,2}$.

Expanding the gauge field functions $\omega _{j}$ to order $r^{2}$ has therefore only introduced one arbitrary parameter,
namely $b_{1}$.
However, it is expected that $N-1$ independent parameters will be required to describe the $N-1$ independent functions
$\omega _{j}(r)$.
Therefore, we must work to higher order in $r$ in order to introduce more arbitrary parameters.

Considering the next order in the Yang-Mills equations (\ref{eq:YMe}), and setting
${\bmath {\omega }}_{3}=\left( \omega _{1,3}, \ldots , \omega _{N-1,3} \right) ^{T}$,
we find
\begin{equation}
{\cal {M}}_{N-1} {\bmath {\omega }}_{3} = 6{\bmath {\omega }}_{3},
\end{equation}
so that we may set
\begin{equation}
{\bmath {\omega }}_{3} = b_{2}{\bmath {v}}_{2},
\end{equation}
where $b_{2}$ is an arbitrary constant and ${\bmath {v}}_{2}$ is an eigenvector of ${\cal {M}}_{N-1}$ with eigenvalue
$6$ ($k=2$ in (\ref{eq:eigenvalues})).
The Einstein equations (\ref{eq:Ee}) are then used to determine $m_{4}$ and $S_{3}$ (which
will also depend on the cosmological constant $\Lambda $).

Since we require $N-1$ arbitrary parameters for the $N-1$ independent functions $\omega _{j}(r)$,
the above analysis therefore suggests that we need to expand the $\omega _{j}(r)$ up to $r^{N+1}$ in order
to have $N-1$ arbitrary parameters in the expansion.
This turns out to be the case, and a detailed proof will be given elsewhere \cite{BW}.
Determining the ${\bmath {\omega }}_{k} = \left( \omega _{1,k}, \ldots , \omega _{N-1,k} \right) ^{T}$
for $k>3$ is slightly more complicated than for $k=2,3$ as outlined above.
Examining the Yang-Mills equation (\ref{eq:YMe}) to $k$th order, we find an equation for the
${\bmath {\omega }}_{k+1}$ of the following form
\begin{equation}
\left[ {\cal {M}}_{N-1} - k\left( k + 1 \right) \right] {\bmath {\omega }}_{k+1} = {\bmath {c}}_{k+1}
\label{eq:genomegak}
\end{equation}
where ${\bmath {c}}_{k+1}$ is a complicated vector depending on
${\bmath {\omega }}_{1},\ldots ,{\bmath {\omega }}_{k}$ and $m_{3},\ldots ,m_{k}$.
For $\Lambda = 0$, the form of ${\bmath {c}}_{k+1}$ is given explicitly in \cite{Kunzle1};
when $\Lambda < 0$ there are minor modifications which we do not write here (they will be given in \cite{BW}).
Since, in later sections, we present solutions just for ${\mathfrak {su}}(3)$ and ${\mathfrak {su}}(4)$
EYM, we will not need the exact form of the ${\bmath {c}}_{k+1}$.
If ${\bmath {v}}_{k}$ is an eigenvector of ${\cal {M}}_{N-1}$ with eigenvalue ${\cal {E}}_{k}$ (\ref{eq:eigenvalues}),
we can solve equation (\ref{eq:genomegak}) for ${\bmath {\omega }}_{k+1}$:
\begin{equation}
{\bmath {\omega }}_{k+1} = b_{k} {\bmath {v}}_{k} + {\bmath {u}}_{k+1},
\end{equation}
where ${\bmath {u}}_{k+1}$ is a particular solution of (\ref{eq:genomegak}) chosen by requiring that
${\bmath {u}}_{k+1}$ is a linear combination of ${\bmath {v}}_{1},\ldots , {\bmath {v}}_{k-1}$.
It is proven in \cite{Kunzle1} that there is a unique solution of (\ref{eq:genomegak}) subject to
this constraint, in the $\Lambda = 0$ case.
This can be extended to $\Lambda <0$, but we do not present the lengthy details here \cite{BW}.

The upshot of all this analysis is that the expansion of the fields, near the origin, is written as follows
(where ${\bmath {\omega }} = \left( \omega _{1}, \ldots , \omega _{N-1} \right) ^{T}$):
\begin{eqnarray}
m(r) & = & m_{3}r^{3} + O(r^{4});
\nonumber \\
S(r) & = & S_{0} + O(r^{2});
\nonumber \\
{\bmath {\omega }}(r) & = & {\bmath {\omega }}_{0} +
\sum _{k=1}^{N-1} b_{k} {\bmath {v}}_{k} r^{k+1} + O(r^{N+1}),
\label{eq:origin}
\end{eqnarray}
where
\begin{equation}
{\bmath {\omega }}_{0} = \left( {\sqrt {N-1}}, {\sqrt {2(N-2)}}, \ldots , {\sqrt {(N-1)}} \right) ^{T}.
\end{equation}
The expansions (\ref{eq:origin}) give the field variables in terms of the $N-1$ parameters $b_{1},\ldots , b_{N-1}$
and are those which are used in the numerical integration of the field equations in the next section.

\section{Solutions}
\label{sec:solutions}

The field equations (\ref{eq:YMe},\ref{eq:Ee}) have the following trivial solutions.
Setting $\omega _{j}(r) \equiv \pm {\sqrt {j(N-j)}}$ for all $j$ gives the Schwarzschild-adS black hole
with $m(r) =M= {\mbox {constant}}$ (which can be set to zero to give pure adS space).
Setting $\omega _{j}(r) \equiv 0 $ for all $j$ gives the Reissner-Nordstr\"om-adS black hole with
metric function
\begin{equation}
\mu (r) = 1 - \frac {2M}{r} + \frac {Q}{r^{2}} - \frac {\Lambda r^{2}}{3},
\end{equation}
and magnetic charge $Q$ given by
\begin{equation}
Q^{2} = \frac {1}{6} N \left( N + 1\right) \left( N - 1 \right) .
\end{equation}
There is an additional special class of solutions, given by setting
\begin{equation}
\omega _{j}(r) =\pm  {\sqrt {j(N-j)}} \, \omega (r) \qquad \forall j=1,\ldots ,N-1.
\label{eq:embeddedsu2}
\end{equation}
In this case, we follow \cite{Kunzle1} and define
\begin{equation}
\lambda _{N} = {\sqrt {\frac {1}{6} N \left( N -1 \right) \left( N +1  \right) }},
\label{eq:lambdaN}
\end{equation}
and then rescale the field variables as follows:
\begin{eqnarray}
R & = & \lambda _{N}^{-1} r; \qquad
{\tilde {\Lambda }} = \lambda _{N}^{2}\Lambda ; \nonumber \\
{\tilde {m}}(R) & = & \lambda _{N}^{-1} m(r); \qquad
{\tilde {S}}(R) = S(r); \nonumber \\
{\tilde {\omega }}(R) & = & \omega (r).
\label{eq:scaling}
\end{eqnarray}
Note that we rescale the cosmological constant $\Lambda $ (this is not necessary in \cite{Kunzle1} as
there $\Lambda = 0$).
The field equations satisfied by ${\tilde {m}}(R)$, ${\tilde {S}}(R)$ and ${\tilde {\omega }}(R)$ are then
\begin{eqnarray}
\frac {d{\tilde {m}}}{dR} & = &
\left[ \mu {\tilde {G}} + R^{2} {\tilde {p}}_{\theta } \right] ; \qquad
\frac {1}{{\tilde {S}}} \frac {d{\tilde {S}}}{dR}  =
- \frac {2{\tilde {G}}}{R} ;
\nonumber \\
0 & = & R^{2} \mu \frac {d^{2}{\tilde {\omega }}}{dR^{2}} +
\left[ 2{\tilde {m}} - 2R^{3} {\tilde {p}}_{\theta } - \frac {2{\tilde {\Lambda }}R^{3}}{3}
\right] \frac {d{\tilde {\omega }}}{dR}
\nonumber \\ & &
+ \left[ 1 - {\tilde {\omega }}^{2} \right] {\tilde {\omega }},
\label{eq:su2equations}
\end{eqnarray}
where we now have
\begin{equation}
\mu  =  1 - \frac {2{\tilde {m}}}{R} - \frac {{\tilde {\Lambda }}R^{2}}{3},
\end{equation}
and
\begin{equation}
{\tilde {G}} = \left( \frac {d{\tilde {\omega }}}{dR} \right) ^{2} , \qquad
{\tilde {p}}_{\theta } = \frac {1}{2R^{2}} \left( 1 - {\tilde {\omega }}^{2} \right) ^{2}.
\end{equation}
The equations (\ref{eq:su2equations}) are precisely the ${\mathfrak {su}}(2)$ EYM field equations.
Furthermore, the boundary conditions (\ref{eq:horizon},\ref{eq:infinity},\ref{eq:origin}) also become
those for the ${\mathfrak {su}}(2)$ case.
This is straightforward to see for the boundary conditions at the horizon (\ref{eq:horizon}) or at
infinity (\ref{eq:infinity}).
At the origin (\ref{eq:origin}), the ${\mathfrak {su}}(2)$ embedded solutions are given by $b_{1}\neq 0$,
but $b_{2}=\ldots = b_{N-1}=0$.
Therefore any ${\mathfrak {su}}(2)$, asymptotically adS, EYM soliton or black hole solution can be embedded into
${\mathfrak {su}}(N)$ EYM to give another asymptotically adS soliton or black hole.
We will see later in this section how the embedded ${\mathfrak {su}}(2)$ solutions fit in the solution
spaces for larger $N$.

To find genuinely ${\mathfrak {su}}(N)$ solutions,
the field equations (\ref{eq:YMe},\ref{eq:Ee}) are integrated numerically using standard `shooting' techniques
\cite{NR}.
The equation for $S(r)$ decouples from the other Einstein equation and the Yang-Mills equations so can be integrated
separately if required.
For ${\mathfrak {su}}({N})$ solutions, we therefore have $N$ coupled ordinary differential equations to integrate
($N-1$ Yang-Mills equations and one Einstein equation).
For black holes, we start integrating just outside the event horizon, using as our shooting parameters
the $N$ variables $\omega _{j}(r_{h})$, $r_{h}$,
subject to the weak constraint (\ref{eq:constraint}).
For solitons, we start integrating close to the origin, using as our shooting parameters the $(N-1)$ variables
$b_{j}$ (\ref{eq:origin}).
In the soliton case, there are no {\em {a priori}} bounds on the parameters $b_{j}$.
In either case, the field equations are then integrated outwards in the radial co-ordinate $r$ until either the
field variables start to
diverge or they have converged to the asymptotic form at infinity.

We now turn to a detailed discussion of the solutions we find.
As well as presenting some examples of solutions, our particular focus in this section will be
the structure of the space of solutions, as a subset of the phase space of parameters
characterizing the solutions.
We will examine the solution spaces in detail for ${\mathfrak {su}}(3)$ and ${\mathfrak {su}}(4)$ solitons
and black holes,
focusing on the numbers of zeros of the gauge field functions.
The solution spaces we present may not necessarily be complete, as our approach has been to scan the
parameter space using a grid.
It is therefore possible that solutions in which the gauge field functions have different numbers of zeros
exist between the points of our grid.
However, our figures will reveal the key features of the solution spaces.
Of particular interest will be the existence of solutions where all the gauge field functions have
no zeros.

\subsection{${\mathfrak {su}}(2)$ solutions}
\label{sec:su2}

We begin by reviewing the phase space of ${\mathfrak {su}}(2)$ solutions, in which
case we have a single gauge field function $\omega (r)$.
Many of the properties we find in the phase space of solutions for ${\mathfrak {su}}(N)$, $N>2$ are
also seen in the ${\mathfrak {su}}(2)$ case and so it is informative to
examine this simpler situation first.

\subsubsection{${\mathfrak {su}}(2)$ solitons}
\label{sec:su2solitons}

\begin{figure}
\begin{center}
\includegraphics[width=6.5cm,angle=270]{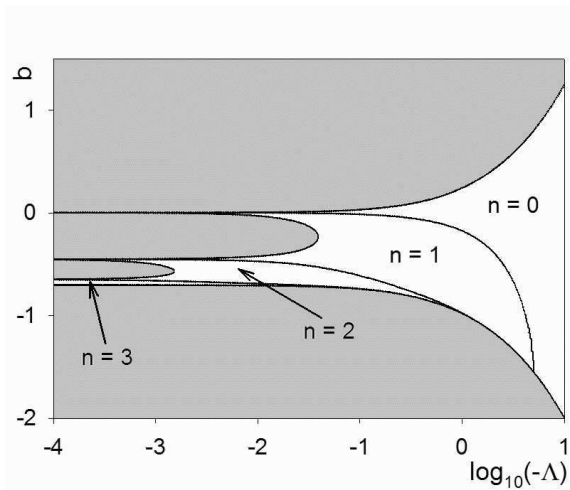}
\end{center}
\caption{Phase space of soliton solutions of ${\mathfrak {su}}(2)$ EYM.
The shaded area denotes those pairs $(\Lambda, b)$ (where $b$ is the shooting parameter giving
the form of the gauge field function $\omega $ near the origin) for which no regular solution is found.
The unshaded regions correspond to regular solutions, the number of nodes $n$ of the gauge field function
$\omega $ being indicated for each region.
For values of $b$ just below the region in which $n=3$ we found solutions for which $n=4$, but the latter
region is too small to be seen on the graph.}
\label{fig:su2solitons}
\end{figure}

Near the origin, one parameter, $b$, is required, and the expansion (\ref{eq:origin}) reduces to
\begin{equation}
\omega (r) = 1 + br^{2} + O\left( r^{3} \right) .
\label{eq:su2solitonorigin}
\end{equation}
For solitons, the phase space has been studied in detail by \cite{BML}.
We have verified their results and the phase space is shown in figure \ref{fig:su2solitons}
(note that our parameter $b$ in (\ref{eq:su2solitonorigin}) is equal to $-b$ in \cite{BML}).
The phase space is parameterized by just two quantities: the cosmological constant $\Lambda $ and the shooting parameter
$b$ (\ref{eq:su2solitonorigin}).
The shaded regions in figure \ref{fig:su2solitons} indicate those values of the parameters for which we were
unable to find a regular solution all the way out to infinity.
Where we do find solutions, they occur in open subsets of the plane.
We label these open sets by $n$, the number of zeros of the single gauge function $\omega $.
We draw the reader's attention to the following particular features of the soliton phase space:
\begin{enumerate}
\item
The number of zeros of the gauge field function increases as $\left| \Lambda \right| $ decreases or $b$ decreases.
\item
Solutions in which $\omega $ has no zeros occur for sufficiently large $\left| \Lambda \right| $.
\item
As $\left| \Lambda \right| $ decreases, we find fewer solutions. The phase space breaks up into smaller and smaller
regions.
In the limit $\Lambda \rightarrow 0$, we are left with solutions just at discrete points, which are the
Bartnik-McKinnon solitons in asymptotically flat space \cite{Bartnik}.
\end{enumerate}
Note that the solution with $b=0$ exists for all $\Lambda $, and simply corresponds to pure adS, with
$\omega (r) \equiv 1$.

\subsubsection{${\mathfrak {su}}(2)$ black holes}
\label{sec:su2bh}

We next turn to the phase space of ${\mathfrak {su}}(2)$ black hole solutions.
There are now three parameters describing the solutions, $r_{h}$, $\Lambda $ and $\omega (r_{h})$.
In order to plot two-dimensional figures, we fix either $r_{h}$ or $\Lambda $ and vary the other two quantities.
For ${\mathfrak {su}}(2)$ black holes, the constraint (\ref{eq:constraint}) on the value of the gauge field
function at the event horizon reads
\begin{equation}
\left( \omega (r_{h}) ^{2} - 1 \right) ^{2} < r_{h}^{2} \left( 1 - \Lambda r_{h}^{2} \right) .
\label{eq:su2bhconstraint}
\end{equation}
Whether we are varying $r_{h}$ or $\Lambda $, we perform a scan over all values of $\omega _{h}$ which
satisfy (\ref{eq:su2bhconstraint}).

\begin{figure}
\begin{center}
\includegraphics[width=7.5cm,angle=270]{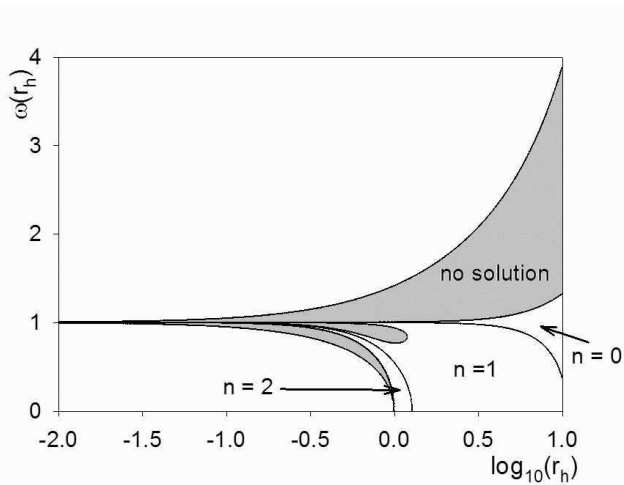}
\end{center}
\caption{The space of ${\mathfrak {su}}(2)$ black hole solutions when $\Lambda = -0.01$, for varying $r_{h}$.
The shaded region indicates values of the gauge field function $\omega (r_{h})$ at the event horizon
for which the constraint (\ref{eq:su2bhconstraint}) is satisfied, but for which we find no well-behaved black
hole solution.  The number of zeros $n$ of the gauge field function $\omega $ are indicated in those regions
of the phase space where we find black hole solutions. Elsewhere on the diagram, the constraint
(\ref{eq:su2bhconstraint}) is not satisfied.
Between the region where $n=2$ and the shaded region we find black hole solutions with $n=3$, $4$ and $5$,
but these regions are too small to indicate on the graph.}
\label{fig:su2bh1}
\end{figure}

Firstly, we show in figure \ref{fig:su2bh1} the space of black hole solutions for fixed $\Lambda = -0.01$ and
varying event horizon radius $r_{h}$.
The outermost curves in figure \ref{fig:su2bh1} are where the inequality (\ref{eq:su2bhconstraint})
is saturated.
Immediately inside these curves we have a shaded region, which represents values of $\left( r_{h}, \omega (r_{h}) \right) $
for which the constraint (\ref{eq:su2bhconstraint}) is satisfied, but for which we are unable to find
black hole solutions which remain regular all the way out to infinity.
As with the solitons, where we do find solutions, we indicate in figure \ref{fig:su2bh1} the number of
zeros of the gauge field function $\omega (r)$.
The solution for which $\omega (r_{h})=1$ is simply the Schwarzschild-adS black hole, while that for $\omega (r_{h})=0$
is the magnetically charged Reissner-Nordstr\"om-adS black hole, as described above.
The following key features are apparent from figure \ref{fig:su2bh1}:
\begin{enumerate}
\item
We find solutions in which the gauge field function has more zeros as we decrease $r_{h}$ or $\omega (r_{h})$.
\item
As $r_{h}\rightarrow 0$, the constraint (\ref{eq:su2bhconstraint}) implies that $\omega (r_{h}) \rightarrow 1$,
as can be seen in figure \ref{fig:su2bh1}.
This is because the black hole solutions become solitons in this limit, and, for solitons, we have $\omega (0)=1$
(\ref{eq:su2solitonorigin}).
However, as can be seen in figure \ref{fig:su2solitons}, for this value of $\Lambda$, there are different
soliton solutions, with $\omega $ having different numbers of zeros.
This feature is not readily apparent from figure \ref{fig:su2bh1}.
\item
The phase space of solutions breaks up into smaller regions as $r_{h}$ decreases.
\end{enumerate}
We find similar behaviour on varying $r_{h}$ for different values of $\Lambda $.

\begin{figure}
\begin{center}
\includegraphics[width=6.2cm,angle=270]{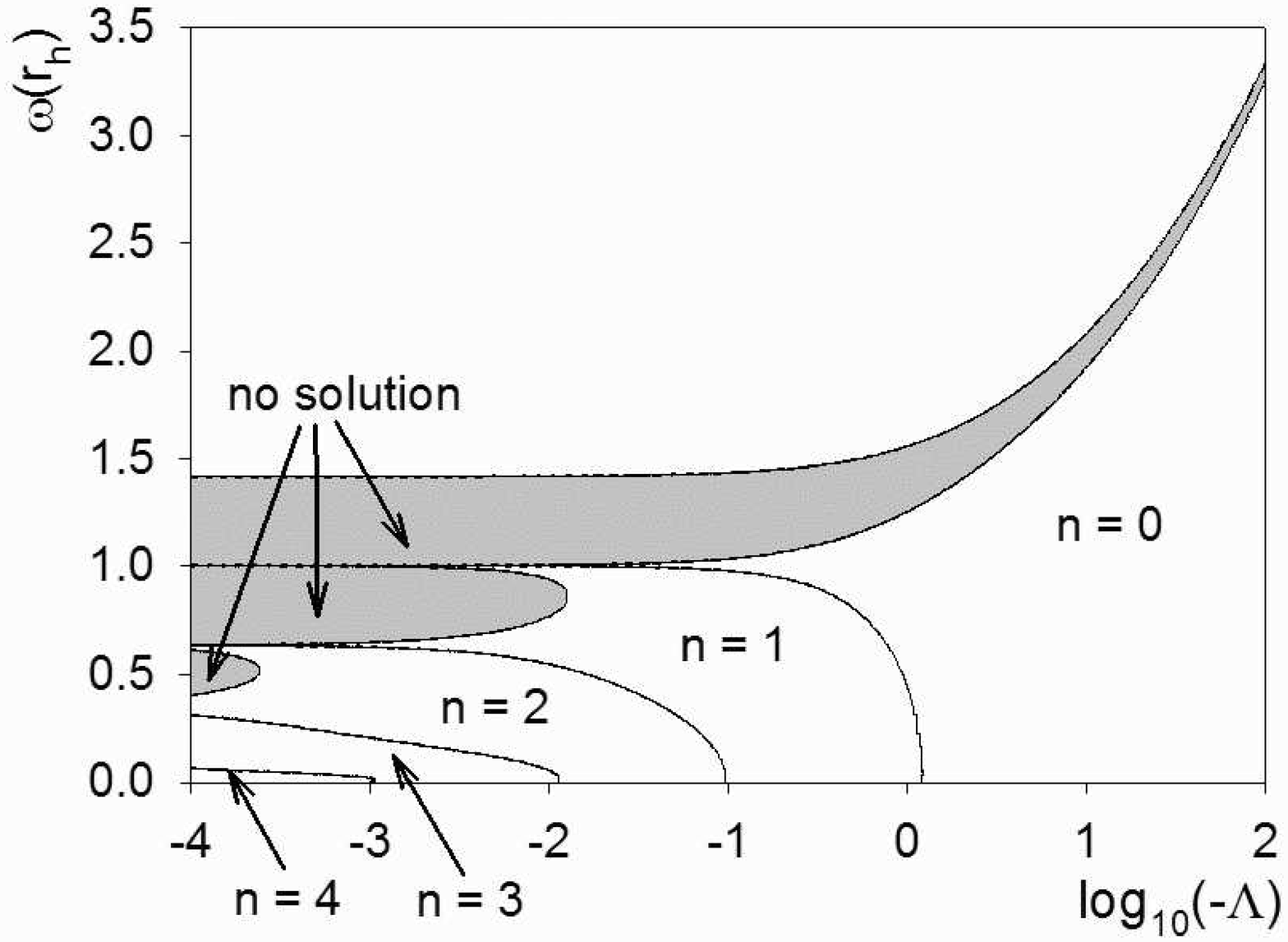}
\end{center}
\caption{Phase space of ${\mathfrak {su}}(2)$ black holes with $r_{h}=1$ and varying $\Lambda $.
The shaded region indicates values of the gauge field function $\omega (r_{h})$ at the event horizon
for which the constraint (\ref{eq:su2bhconstraint}) is satisfied, but for which we find no well-behaved black
hole solution.  The number of zeros $n$ of the gauge field function $\omega $ are indicated in those regions
of the phase space where we find black hole solutions. Elsewhere on the diagram, the constraint
(\ref{eq:su2bhconstraint}) is not satisfied.
As well as the regions where $n=0,\ldots,4$ as marked on the diagram, we find a small region in
the bottom left of the plot where $n=5$. This region is too small to indicate on the current figure,
but can be seen in figure \ref{fig:su2bh3}.}
\label{fig:su2bh2}
\end{figure}

\begin{figure}
\begin{center}
\includegraphics[width=6.2cm,angle=270]{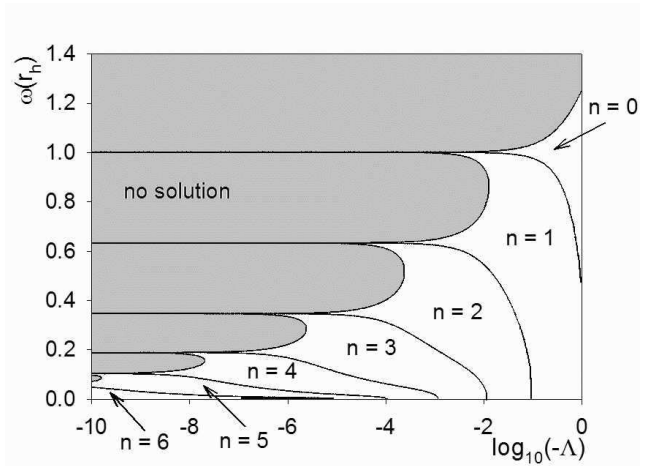}
\end{center}
\caption{Close-up of the phase space of ${\mathfrak {su}}(2)$ black holes with $r_{h}=1$ and smaller values
of $\Lambda $.
In the bottom left of the plot there is a small region of solutions for which $n=7$, but the region
is too small to be visible.}
\label{fig:su2bh3}
\end{figure}

We now fix the event horizon radius $r_{h}=1$ and vary the cosmological constant $\Lambda $.
The solution space in this case is shown in figure \ref{fig:su2bh2}, with a close-up for smaller values of
$\left| \Lambda \right| $ in figure \ref{fig:su2bh3}.
Once again, in figures \ref{fig:su2bh2} and \ref{fig:su2bh3} we have shaded those regions where the constraint
(\ref{eq:su2bhconstraint}) is satisfied, but no regular black hole solutions could be found.
Where we do find solutions, the number of zeros of the gauge field function $\omega (r)$ is indicated in the figures.
Similar behaviour is observed as in the soliton case (figure \ref{fig:su2solitons}), namely:
\begin{enumerate}
\item
The number of zeros of the gauge field function increases as $\left| \Lambda \right| $ or $\omega (r_{h})$ decreases.
\item
As $\Lambda \rightarrow 0$, the phase space breaks up into discrete points, which correspond to the asymptotically
flat `colored' ${\mathfrak {su}}(2)$ black holes \cite{Bizon}.
\item
For sufficiently large $\left| \Lambda \right| $, we find solutions in which the gauge field function has no zeros.
\end{enumerate}

\subsection{Black holes}
\label{sec:BH}

We now turn to solutions of the ${\mathfrak {su}}(N)$ EYM field equations with $N>2$, considering firstly black holes and
then solitons.
Many of the features of the ${\mathfrak {su}}(2)$ solutions outlined in the previous section will be
replicated for larger $N$.

\subsubsection{${\mathfrak {su}}(3)$ black holes}
\label{sec:su3bh}

\begin{figure}
\begin{center}
\includegraphics[width=6.2cm,angle=270]{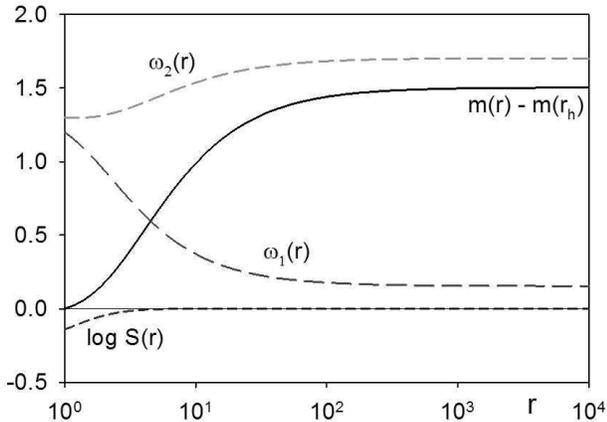}
\end{center}
\caption{Typical ${\mathfrak {su}}(3)$ black hole solution, with $r_{h}=1$, $\Lambda = -1$,
$\omega _{1}(r_{h}) = 1.2$ and $\omega _{2}(r_{h}) = 1.3$.
In this example, both gauge field functions have no zeros.}
\label{fig:su3bhex1}
\end{figure}

\begin{figure}
\begin{center}
\includegraphics[width=6.5cm,angle=270]{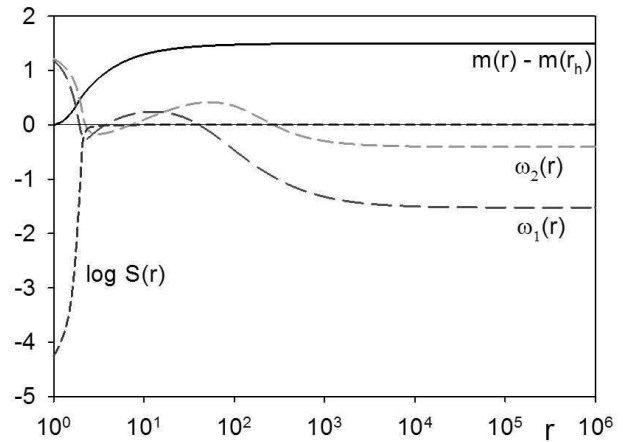}
\end{center}
\caption{Example of an ${\mathfrak {su}}(3)$ black hole solution, with $r_{h}=1$,
$\Lambda = -0.0001$, $\omega _{1}(r_{h})=1.184$ and $\omega _{2}(r_{h})=1.216$.
In this case, both gauge field functions have three zeros.}
\label{fig:su3bhex2}
\end{figure}

\begin{figure}
\begin{center}
\includegraphics[width=6cm,angle=270]{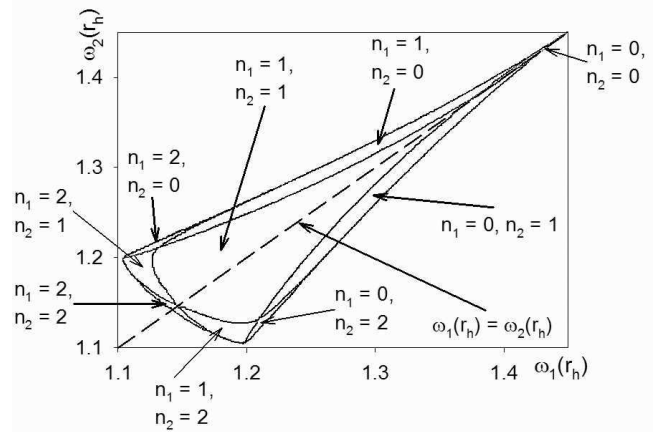}
\end{center}
\caption{Solution space for ${\mathfrak {su}}(3)$ black holes with $r_{h}=1$ and $\Lambda = -0.1$.
The numbers of zeros of the gauge field functions for the various regions of the solution space are shown.
For other values of $\omega _{1}(r_{h})$, $\omega _{2}(r_{h})$ we find no solutions.
There is a very small region containing solutions in which both gauge field functions have no zeros,
in the top-right-hand corner of the plot.}
\label{fig:su3bh3}
\end{figure}

\begin{figure}
\begin{center}
\includegraphics[width=6.5cm,angle=270]{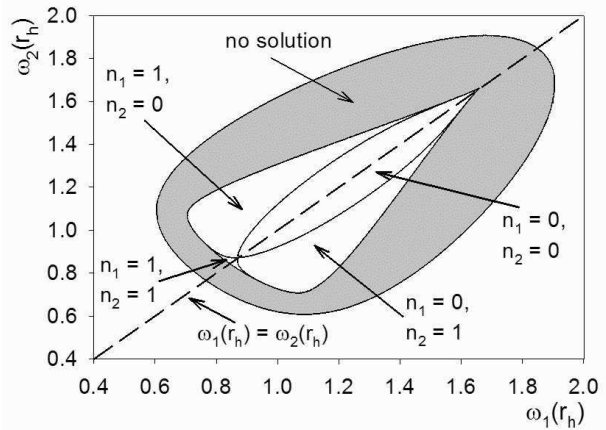}
\end{center}
\caption{Solution space for ${\mathfrak {su}}(3)$ black holes with $r_{h}=1$ and $\Lambda = -1$.
The shaded region indicates where the constraint (\ref{eq:su3bhconstraint}) is satisfied but we do
not find black hole solutions.
Outside the shaded region the constraint (\ref{eq:su3bhconstraint}) does not hold.
Where there are solutions, we have indicated the numbers of zeros of the gauge field functions
within the different regions.
For this value of $\Lambda $ there is a large region in which both gauge field functions have no zeros.}
\label{fig:su3bh2}
\end{figure}

For ${\mathfrak {su}}(3)$ EYM, there are two gauge field functions $\omega _{1}(r)$ and $\omega _{2}(r)$,
and therefore four parameters describing black hole solutions: $r_{h}$, $\Lambda $, $\omega _{1}(r_{h})$
and $\omega _{2}(r_{h})$.
Using the symmetry of the field equations (\ref{eq:omegaswap}),
we set $\omega _{1}(r_{h}),\omega _{2}(r_{h})>0$ without loss of generality.
The constraint (\ref{eq:constraint}) on the values of the gauge field functions at the horizon becomes, in this
case:
\begin{eqnarray}
& & \left[ \omega _{1}(r_{h})^{2} - 2 \right] ^{2}
+ \left[ \omega _{1}(r_{h})^{2} - \omega _{2}(r_{h})^{2} \right] ^{2}
\nonumber \\ & & \qquad
+ \left[ 2 - \omega _{2}(r_{h})^{2} \right] ^{2}
\nonumber \\ & & \qquad
<
2 r_{h}^{2} \left( 1 - \Lambda r_{h}^{2} \right) .
\label{eq:su3bhconstraint}
\end{eqnarray}
Two typical black hole solutions are shown in figures \ref{fig:su3bhex1} and \ref{fig:su3bhex2}.
The metric functions behave in a very similar way to the ${\mathfrak {su}}(2)$ solutions
\cite{ew1,Bjoraker}, smoothly interpolating
between their values at the horizon and at infinity.
We note that $S(r)$ in particular converges very rapidly to $1$ as $r\rightarrow \infty $.
In figure \ref{fig:su3bhex1}, we show an example of a black hole solution in which both gauge field functions
have no zeros.
We note that both gauge field functions are monotonic, however, one is monotonically increasing and the other
monotonically decreasing.
In our second example (figure \ref{fig:su3bhex2}) both gauge field functions have three zeros.
Although, in both our examples the two gauge field functions have the same number of zeros, we
also find solutions where the two gauge field functions have different numbers of zeros (see figures
\ref{fig:su3bh3} and \ref{fig:su3bh2}).

We now examine the space of black hole solutions.
Since we have four parameters, in order to produce two-dimensional figures, we need to fix two parameters in each case.
We find that varying the event horizon radius produces similar behaviour to the ${\mathfrak {su}}(2)$ case, so
for the remainder of this section we fix $r_{h}=1$ and consider the phase space for different, fixed values of $\Lambda $,
scanning all values of $\omega _{1}(r_{h})$, $\omega _{2}(r_{h})$ such that the constraint (\ref{eq:su3bhconstraint})
is satisfied.
From the discussion at the beginning of section \ref{sec:solutions}, we have embedded ${\mathfrak {su}}(2)$
black hole solutions when, from (\ref{eq:embeddedsu2}):
\begin{equation}
\omega _{1}(r)={\sqrt {2}}\omega (r)=\omega _{2}(r)
\end{equation}
which occurs when $\omega _{1}(r_{h})=\omega _{2}(r_{h})$.

\begin{figure}
\begin{center}
\includegraphics[width=6.2cm,angle=270]{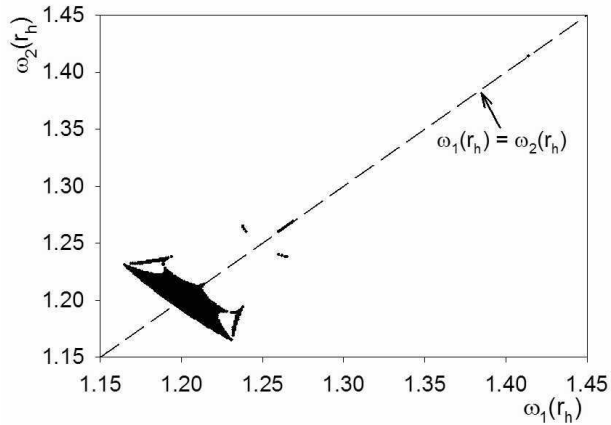}
\end{center}
\caption{Solution space for ${\mathfrak {su}}(3)$ black holes with $r_{h}=1$ and $\Lambda = -0.0001$.
The dashed line indicates where $\omega _{1}(r_{h})=\omega _{2}(r_{h})$, along which lie the
embedded ${\mathfrak {su}}(2)$ solutions.
The black regions indicate where we have regular black hole solutions; elsewhere we find no solutions.
In this case we find a wide variety of numbers of zeros of the gauge field functions, and so do not indicate
all the different possibilities.
For these values of $\Lambda $ and $r_{h}$ we find no solutions for which the gauge field functions have no zeros.
The key feature in this figure is the fragmentation of the solution space and the fact that there are comparatively few
solutions.}
\label{fig:su3bh1}
\end{figure}

\begin{figure}
\begin{center}
\includegraphics[width=6.5cm,angle=270]{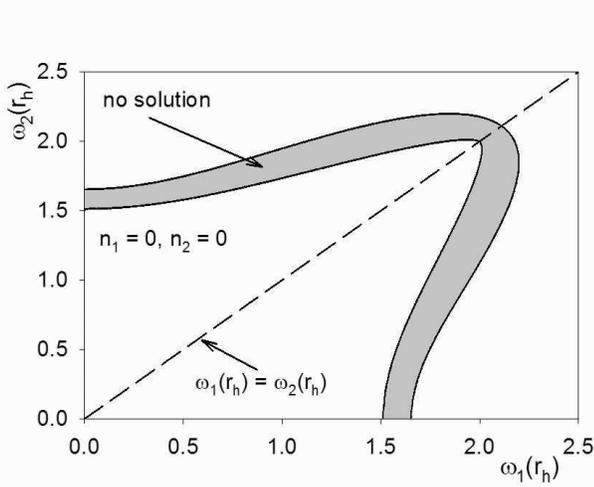}
\end{center}
\caption{Solution space for ${\mathfrak {su}}(3)$ black holes with $r_{h}=1$ and $\Lambda = -5$.
It can be seen that for the vast majority of the phase space for which the constraint (\ref{eq:su3bhconstraint})
is satisfied, we have black hole solutions in which both gauge field functions have no zeros.}
\label{fig:su3bh4}
\end{figure}

In figures \ref{fig:su3bh1}-\ref{fig:su3bh4} we plot the phase space of solutions for fixed
event horizon radius $r_{h}=1$ and varying cosmological constant $\Lambda = -0.0001$, $-0.1$, $-1$ and $-5$ respectively.
In each of figures \ref{fig:su3bh1}-\ref{fig:su3bh4} we plot the dashed line $\omega _{1}(r_{h})=\omega _{2}(r_{h})$,
along which lie the embedded ${\mathfrak {su}}(2)$ black holes.
It is seen in all these figures that the solution space is symmetric about this line, as would be expected from
the symmetry (\ref{eq:Nswap}) of the field equations.

As in the ${\mathfrak {su}}(2)$ case, for small values of $\Lambda $ (see figure \ref{fig:su3bh1})
the solution space fragments and we find very few solutions.
The values of $\left( \omega _{1}(r_{h}), \omega _{2}(r_{h}) \right) $ for which we find regular black hole
solutions are indicated by black dots in figure
\ref{fig:su3bh1}.
Above the main group of solutions, there can clearly be seen a couple of smaller regions of solutions.
There is also a small region centered on and very close to the dashed line at about $\omega _{1}(r_{h}) \sim 1.27$,
and the Schwarzschild-adS solution at $\omega _{1}(r_{h})=\omega _{2}(r_{h})={\sqrt {2}}$.
For this value of $\Lambda$, we find very complicated behaviour in the numbers of zeros
$(n_{1},n_{2})$ of the gauge field functions $\omega _{1}(r)$, $\omega _{2}(r)$ respectively.
We have found at least fourteen different combinations of the numbers of zeros of the gauge field functions,
some of which occur only in very small regions of the parameter space.
This behaviour is too complicated to depict accurately in figure \ref{fig:su3bh1}.
The numbers of zeros of the gauge field functions vary between 1 and 4 (we find no solutions in which either
gauge field function has no zeros).
We stress that the gauge field functions do not have to have the same numbers of zeros, for $\Lambda = -0.0001$
we find that $\left| n_{1}-n_{2} \right| $ varies between 0 and 2.

The solution space is found to be symmetric about the line $\omega _{1}(r_{h})=\omega _{2}(r_{h})$ not only in terms of
where we find solutions, but also in terms of the numbers of zeros of the gauge field functions.
To state this precisely, suppose that at the point $\omega _{1}(r_{h})=a_{1}$, $\omega _{2}(r_{h})=a_{2}$
we find a black hole solution in which $\omega _{1}(r)$ has $n_{1}$ zeros and $\omega _{2}(r)$ has $n_{2}$ zeros.
Then, at the point $\omega _{1}(r) = a_{2}$, $\omega _{2}(r) = a_{1}$, we find a black hole solution
in which $\omega _{1}(r)$ has $n_{2}$ zeros and $\omega _{1}(r)$ has $n_{1}$ zeros.
This is clearly seen in figures \ref{fig:su3bh3} and \ref{fig:su3bh2}, and follows from the symmetry
(\ref{eq:Nswap}) of the field equations.

As we increase $\left| \Lambda \right| $, we find
(see figures \ref{fig:su3bh3}-\ref{fig:su3bh4})
that the solution space expands as a proportion of the
space of values of $\omega _{1}(r_{h})$, $\omega _{2}(r_{h})$ satisfying the constraint (\ref{eq:su3bhconstraint}).
It can also be seen from figures \ref{fig:su3bh3}-\ref{fig:su3bh4} that the number of nodes of the gauge
field functions decreases as $\left| \Lambda \right| $ increases, and that the space of solutions becomes simpler.

For $\Lambda = -0.1$, there is a very small region of the solution space where both gauge field functions have
no zeros.
This region expands as we increase $\left| \Lambda \right| $, until for $\Lambda =-5$, both gauge field
functions have no zeros for all the solutions we find.

\subsubsection{${\mathfrak {su}}(4)$ black holes}
\label{sec:su4bh}

In this case there are three gauge field functions and so the parameter space is five-dimensional.
The constraint (\ref{eq:constraint}) satisfied at the horizon by the gauge field functions now reads:
\begin{eqnarray}
& & \left[ \omega _{1}(r_{h})^{2} - 3 \right] ^{2}
+ \left[ \omega _{2}(r_{h})^{2} - \omega _{1}(r_{h})^{2} -1 \right] ^{2}
\nonumber \\ & & \qquad
+ \left[ \omega _{3}(r_{h})^{2} - \omega _{2}(r_{h})^{2} + 1 \right] ^{2}
+ \left[ 3 - \omega _{3}(r_{h})^{2} \right] ^{2}
\nonumber \\ & & \qquad
<
2 r_{h}^{2} \left( 1 - \Lambda r_{h}^{2} \right) .
\label{eq:su4bhconstraint}
\end{eqnarray}
An example of a typical ${\mathfrak {su}}(4)$ EYM black hole was plotted in \cite{BHW}.
The solutions have the expected features, with the metric functions monotonically interpolating between their
values on the black hole event horizon and at infinity, and the gauge field functions having various
numbers of zeros outside the event horizon before monotonically converging to their values at infinity.

\begin{figure}
\begin{center}
\includegraphics[width=6.5cm,angle=270]{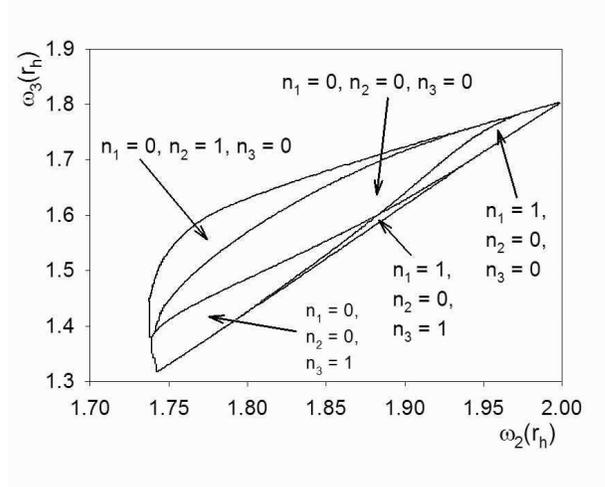}
\end{center}
\caption{Solution space for ${\mathfrak {su}}(4)$ EYM black holes with $r_{h}=1$, $\Lambda = -1$ and
$\omega _{1}(r_{h})=1.6$.
Where there are solutions, the numbers of zeros of the three gauge field functions are indicated for
the relevant regions.
Elsewhere in the figure we find no black hole solutions.
As well as the regions indicated, we also find small regions where the numbers of zeros of the gauge
field functions are $\left( n_{1},n_{2},n_{3}\right) = (1,1,0)$ and $(0,1,1)$.}
\label{fig:su4bh1}
\end{figure}

\begin{figure}
\begin{center}
\includegraphics[width=6.5cm,angle=270]{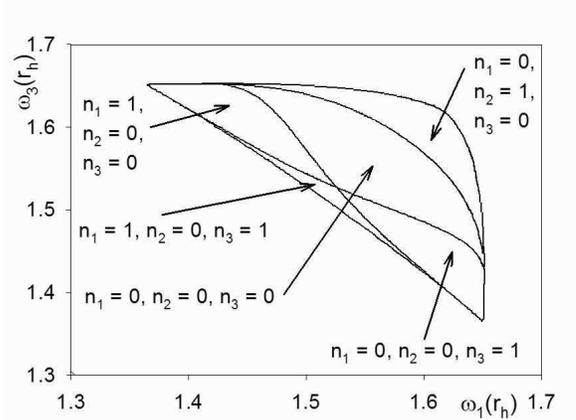}
\end{center}
\caption{Solution space for ${\mathfrak {su}}(4)$ EYM black holes with $r_{h}=1$, $\Lambda = -1$ and
$\omega _{2}(r_{h})=1.8$.
Where there are solutions, the numbers of zeros of the three gauge field functions are indicated for
the relevant regions.
Elsewhere in the figure we find no black hole solutions.
As well as the regions indicated, we also find small regions where the numbers of zeros
of the gauge field functions are $\left( n_{1}, n_{2}, n_{3} \right) = (1,1,0)$ and $(0,1,1)$.}
\label{fig:su4bh2}
\end{figure}

Considering the solution spaces,
to produce a two-dimensional plot, we now have to fix three parameters.
In figures \ref{fig:su4bh1} and \ref{fig:su4bh2} we show examples of the solution space when we fix
$\Lambda $, $r_{h}$ and the value of one of the gauge field functions at the horizon, varying the values
of the other two gauge field functions at the horizon.
In both figures \ref{fig:su4bh1} and \ref{fig:su4bh2} we indicate the numbers of zeros of the three
gauge field functions for the regions where we find black hole solutions.
Elsewhere in these two figures, we do not find black hole solutions.
Now that there are three gauge field functions, it can be seen that the structure of the solution
space is quite complicated (and gets ever more complicated as $\left| \Lambda \right| $ decreases).
However, for the particular values of $\Lambda $ and $r_{h}$ in figures \ref{fig:su4bh1} and
\ref{fig:su4bh2}, it can be seen that there are solutions in which all three gauge field functions
have no zeros.

\begin{figure}
\begin{center}
\includegraphics[width=6.5cm,angle=270]{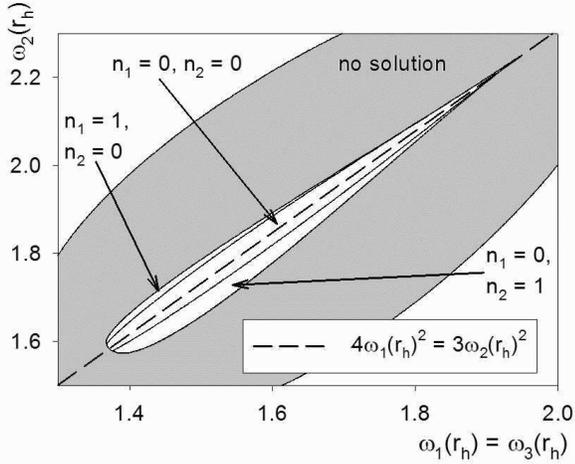}
\end{center}
\caption{Solution space for ${\mathfrak {su}}(4)$ EYM black holes with $\Lambda = -1$, $r_{h}=1$ and
$\omega _{1}(r_{h})=\omega _{3}(r_{h})$.
The shaded region indicates those values of the parameters $\omega _{1}(r_{h})$, $\omega _{2}(r_{h})$
for which the constraint (\ref{eq:su4bhconstraint}) is satisfied, but for which we find no black hole solutions.
Where we do find black hole solutions, the numbers of zeros of the gauge field functions are indicated
(note that $n_{3}=n_{1}$ in this case).
We have also plotted the dashed line $4\omega _{1}(r_{h})^{2}=3\omega _{2}(r_{h})^{2}$, on which
lie embedded ${\mathfrak {su}}(2)$ solutions.
As well as the regions marked, we also find small regions where the numbers of zeros of the gauge field
functions are $n_{1}=n_{3}=2$, $n_{2}=0$ and $n_{1}=n_{3}=0$, $n_{2}=2$.}
\label{fig:su4bh3}
\end{figure}

\begin{figure}
\begin{center}
\includegraphics[width=6cm,angle=270]{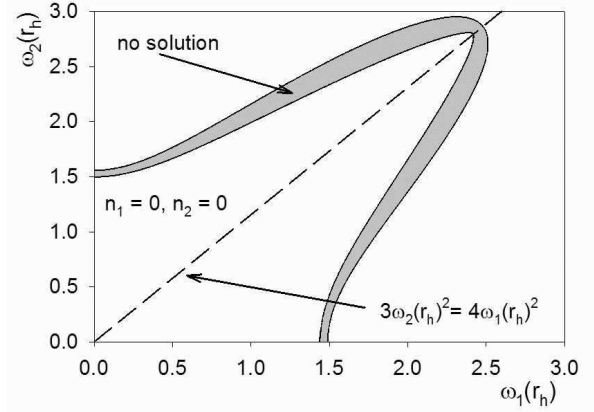}
\end{center}
\caption{Solution space for ${\mathfrak {su}}(4)$ EYM black holes with $\Lambda = -10$, $r_{h}=1$ and
$\omega _{1}(r_{h})=\omega _{3}(r_{h})$.
The shaded region indicates those values of the parameters $\omega _{1}(r_{h})$, $\omega _{2}(r_{h})$
for which the constraint (\ref{eq:su4bhconstraint}) is satisfied, but for which we find no black hole solutions.
In this case, for all the black hole solutions we find, all three gauge field functions have no zeros.
We have also plotted the dashed line $4\omega _{1}(r_{h})^{2}=3\omega _{2}(r_{h})^{2}$, on which
lie embedded ${\mathfrak {su}}(2)$ solutions.}
\label{fig:su4bh4}
\end{figure}

Many of the other features of the phase space observed in the ${\mathfrak {su}}(2)$ and ${\mathfrak {su}}(3)$ cases
are replicated here, namely:
the fragmentation of the solution space as $\left| \Lambda \right| $ decreases;
as $\left| \Lambda \right| $ increases, the proportion of the parameter space for which the constraint
(\ref{eq:su4bhconstraint}) is satisfied and we have black hole solutions increases;
for sufficiently large $\left| \Lambda \right| $, we have solutions in which all gauge field functions have
no zeros.
Figures \ref{fig:su4bh3} and \ref{fig:su4bh4} illustrate these features.
In both figures \ref{fig:su4bh3} and \ref{fig:su4bh4}, we have used the exploited the symmetry (\ref{eq:Nswap})
of the field equations and set $\omega _{1}(r_{h})=\omega _{3}(r_{h})$, although it should be noted, from figures
\ref{fig:su4bh1} and \ref{fig:su4bh2}, that this does not need to hold (that is, although the field equations
have the symmetry (\ref{eq:Nswap}), it is not necessary for the solutions to have this symmetry).
In both figures \ref{fig:su4bh3} and \ref{fig:su4bh4}, we have plotted the line
$4\omega _{1}(r_{h})^{2}=3\omega _{2}(r_{h})^{2}=4\omega _{3}(r_{h})^{2}$, on which lie the embedded
${\mathfrak {su}}(2)$ solutions (\ref{eq:embeddedsu2}).
It can be seen in both figures that the solution space is symmetric about this line, as in the
${\mathfrak {su}}(3)$ case.
Figure \ref{fig:su4bh4} shows that it is still the case that, for sufficiently large $\left| \Lambda \right| $,
all the solutions we find are such that all three gauge field functions have no zeros.

Comparing figures \ref{fig:su3bh2} and \ref{fig:su4bh3}, we see that the proportion of the phase space for which
the constraint (\ref{eq:su4bhconstraint}) is satisfied and we have black hole solutions is rather smaller
than in the ${\mathfrak {su}}(3)$ case.
This can be understood from the scaling (\ref{eq:scaling}) required to embed the ${\mathfrak {su}}(2)$ solutions
into ${\mathfrak {su}}(3)$ EYM.
From (\ref{eq:scaling}), an ${\mathfrak {su}}(2)$ black hole solution with cosmological constant $\Lambda $ and
event horizon radius $r_{h}$ is embedded into ${\mathfrak {su}}(3)$ as a solution with cosmological constant
$4\Lambda $ and event horizon radius $r_{h}/2 $ (since $\lambda _{3} = 2$ (\ref{eq:lambdaN})),
and into ${\mathfrak {su}}(4)$ as a solution with cosmological constant $10\Lambda $ and event horizon
radius $r_{h}/{\sqrt {10}}$ (since $\lambda _{4} = {\sqrt {10}}$ (\ref{eq:lambdaN})).
This scaling means that, for larger $N$,
larger $\Lambda $ values are needed to find the same behaviour as is observed at smaller $\Lambda $ values
in the ${\mathfrak {su}}(2)$ case.

\subsection{Solitons}
\label{sec:solitons}

The behaviour of the gauge field functions near the origin (\ref{eq:origin}) makes finding numerical soliton
solutions of the field equations (\ref{eq:YMe},\ref{eq:Ee}) more complicated than finding black hole solutions.
We define new variables $\beta _{1}(r), \ldots \beta _{N-1}(r)$ which have the following behaviour near the
origin:
\begin{equation}
\beta _{j}(r) = b_{j} r^{j+1} + O \left( r^{j+2} \right) ,
\qquad j = 1, \ldots , N-1;
\label{eq:betaorigin}
\end{equation}
where the $b_{j}$ are the constants in the expansion of ${\bmath {\omega }}(r)$ (\ref{eq:origin}).
Therefore the gauge field functions take the form
\begin{equation}
{\bmath {\omega }}(r)  =  {\bmath {\omega }}_{0} +
\sum _{k=1}^{N-1} \beta_{k}(r) {\bmath {v}}_{k} .
\label{eq:omegabeta}
\end{equation}
For each $N$, we proceed as follows.
Firstly, the normalized eigenvectors ${\bmath {v}}_{k}$ of the matrix ${\cal {M}}_{N-1}$ (\ref{eq:calM})
are calculated.
We then have the $\omega _{j}(r)$ in terms of the $\beta _{k}(r)$ from (\ref{eq:omegabeta}).
The expressions (\ref{eq:omegabeta}) are substituted into the field equations (\ref{eq:YMe},\ref{eq:Ee})
to give differential equations for the $\beta _{k}(r)$.
The Yang-Mills equations for the $\beta _{k}(r)$ will be given explicitly for ${\mathfrak {su}}(3)$ below.
For the Einstein equations, the quantity $G$ (\ref{eq:Gdef}) becomes
\begin{equation}
G =\sum^{N-1}_{k=1}\beta _{k}'^2,
\end{equation}
because we have normalized the eigenvectors ${\bmath {v}}_{k}$.
The quantity $p_{\theta }$ (\ref{eq:ptheta}) takes a complicated form in terms of the $\beta _{k}(r)$
(which we do not write here), but is readily computed in Maple.
Further details of this procedure in the ${\mathfrak {su}}(3)$ and ${\mathfrak {su}}(4)$ cases will be
outlined below.

Many of the features of the solution space for black holes are seen also in the soliton solution spaces.
In particular, the solution space becomes more complicated as $\left| \Lambda \right| $ decreases,
eventually reducing to the asymptotically flat solution space as $\Lambda \rightarrow 0$.
As $\left| \Lambda \right| $ increases, we find more solutions and, for sufficiently large
$\left| \Lambda  \right| $, we find solutions in which all the gauge field functions have no zeros.
In the following subsections, we have focused on the structure of the solution spaces for smaller values
of $\left| \Lambda  \right| $ where there are more features.

\subsubsection{${\mathfrak {su}}(3)$ solitons}
\label{sec:su3solitons}

In ${\mathfrak {su}}(3)$ EYM, the matrix ${\cal {M}}_{N-1}$  (\ref{eq:calM}) with $N=3$ takes the form
\begin{equation}
{\cal {M}}_{2} =
\left(
\begin{array}{cc}
4 & -2 \\
-2 & 4
\end{array}
\right) .
\end{equation}
It is straightforward to confirm that the eigenvalues of ${\cal {M}}_{2}$ are $2$, $6$, with corresponding
normalized eigenvectors
\begin{equation}
{\bmath {v}}_{1} = \frac {1}{{\sqrt {2}}} \left(
\begin{array}{c}
1 \\ 1
\end{array} \right) ;
\qquad
{\bmath {v}}_{2} = \frac {1}{{\sqrt {2}}} \left(
\begin{array}{c}
1 \\ -1
\end{array} \right) .
\end{equation}
As described above, we therefore write the gauge field functions as follows, from (\ref{eq:omegabeta}):
\begin{equation}
\left( \begin{array}{c}
\omega _{1}(r) \\ \omega _{2}(r)
\end{array} \right)
= \left( \begin{array}{c}
{\sqrt {2}} \\ {\sqrt {2}}
\end{array} \right)
+ \frac {1}{{\sqrt {2}}} \left( \begin{array}{c}
\beta _{1}(r) + \beta _{2}(r) \\ \beta _{1}(r) - \beta _{2} (r)
\end{array} \right) .
\label{eq:su3solgauge}
\end{equation}
The Yang-Mills equations (\ref{eq:YMe}) then give the following equations for the $\beta _{k}(r)$:
\begin{eqnarray}
0 & = &
r^{2} \mu \beta _{1}'' +\left( 2m-2r^3 p_{\theta }-\frac {2\Lambda r^3}{3}\right) \beta _{1}'
\nonumber \\ & &
-\frac {1}{4} \left( 2 + \beta _{1} \right) \left( \beta _{1}^{2} + 4\beta _{1} + 7 \beta _{2}^{2} \right) ;
\nonumber \\
0 & = &
r^{2} \mu \beta _{2}'' +\left( 2m-2r^3 p_{\theta }-\frac {2\Lambda r^3}{3}\right) \beta _{2}'
\nonumber \\ & &
-\frac {1}{4} \left( 7\beta _{1}^{2} + 28\beta _{1} + \beta _{2}^{2} + 24 \right) \beta _{2} ;
\label{eq:YMsu3sol}
\end{eqnarray}
where, in this case, the expression for $p_{\theta } $ (\ref{eq:ptheta}) is not too complicated:
\begin{eqnarray}
p_{\theta } & = &
\frac {1}{8r^{4}} \left[
\left( \beta _{1}^{2} + 4\beta _{1} + \beta _{2}^{2} \right) ^{2}
\right. \nonumber \\ & & \left.
+ 48 \beta _{2}^{2} + 48 \beta _{1} \beta _{2}^{2} + 12 \beta _{1}^{2} \beta _{2}^{2}
\right] .
\end{eqnarray}
We then numerically integrate the field equations (\ref{eq:Ee},\ref{eq:YMsu3sol}) with the initial conditions
(\ref{eq:betaorigin}).
The solution space is described by three parameters: $\Lambda $, $b_{1}$ and $b_{2}$.

\begin{figure}
\begin{center}
\includegraphics[width=6cm,angle=270]{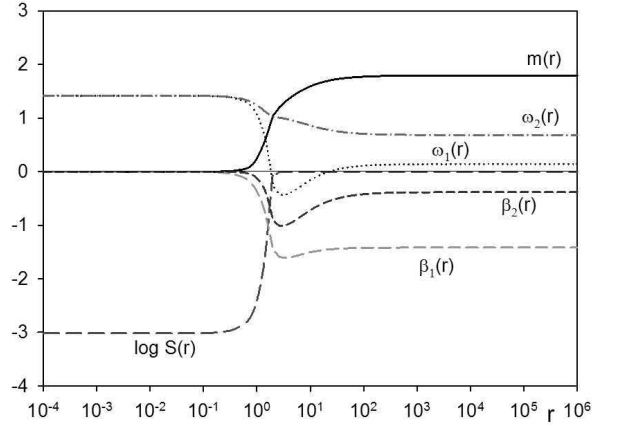}
\end{center}
\caption{Typical ${\mathfrak {su}}(3)$ soliton solution, with $\Lambda = -0.1$, $b_{1} = 0.35$ and $b_{2}=0.115$.
The gauge field function $\omega _{1}(r)$ has two zeros, while $\omega _{2}(r)$ has no zeros.}
\label{fig:su3solitonsex}
\end{figure}

A typical soliton solution is shown in figure \ref{fig:su3solitonsex}.
In figure \ref{fig:su3solitonsex}, we have plotted the auxiliary functions $\beta _{1}(r)$ and $\beta _{2}(r)$
as well as the physical field quantities $m(r)$, $S(r)$, $\omega _{1}(r)$ and $\omega _{2}(r)$.
It can be seen that all variables have the expected behaviour, both near the origin and at infinity.
At infinity, the $\beta _{k}(r)$ functions converge to constant values, which can be arbitrary (since the values
of the gauge field functions $\omega _{j}(r)$ at infinity (\ref{eq:infinity}) are arbitrary).

\begin{figure}
\begin{center}
\includegraphics[width=6cm,angle=270]{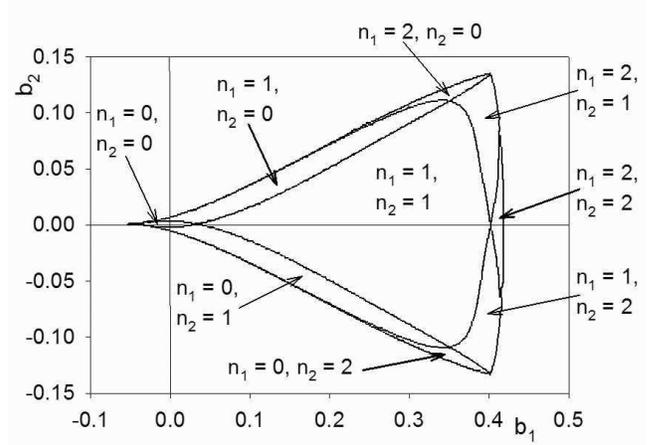}
\end{center}
\caption{Solution space for ${\mathfrak {su}}(3)$ solitons with $\Lambda = -0.1$.
Where we find solutions, the numbers of zeros of the gauge field functions are indicated.
Elsewhere in the parameter space we do not find solutions.
For this value of $\Lambda $, there is a very small region (near the origin) of solutions in which both gauge field
functions have no zeros.
Although they are too small to show on this figure, we also find regions where the numbers of zeros
of the gauge field functions are $n_{1}=3$ (with $n_{2}\in (0,1,2,3)$) or $n_{2}=3$ (with $n_{1}\in (0,1,2,3)$). }
\label{fig:su3solitons1}
\end{figure}

\begin{figure}
\begin{center}
\includegraphics[width=6.5cm,angle=270]{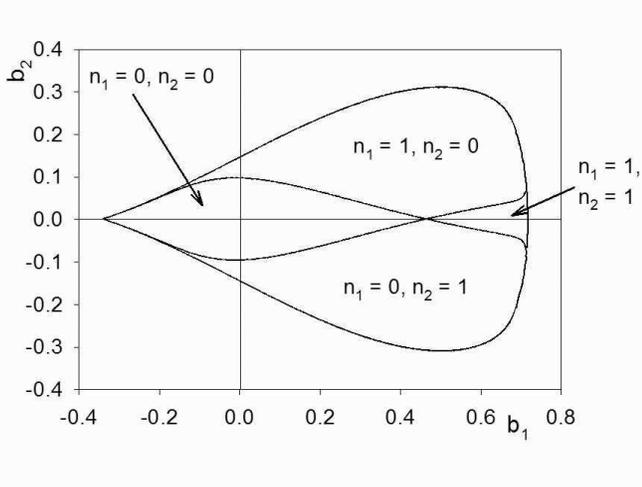}
\end{center}
\caption{Solution space for ${\mathfrak {su}}(3)$ solitons with $\Lambda = -1$.
Where we find solutions, the numbers of zeros of the gauge field functions are indicated.
Elsewhere in the parameter space we do not find solutions.
We now have a much larger region of solutions in which both gauge field functions have no zeros.}
\label{fig:su3solitons2}
\end{figure}

The solution spaces for two particular values of $\Lambda $ are shown in figures \ref{fig:su3solitons1} and
\ref{fig:su3solitons2}.
The origin $b_{1}=0=b_{2}$ corresponds to pure adS space.
In both figures \ref{fig:su3solitons1} and \ref{fig:su3solitons2}, we see that the solution space is symmetric
about the axis $b_{2}=0$.
This is due to the symmetry (\ref{eq:Nswap}) of the field equations, since the mapping $b_{2} \rightarrow -b_{2}$
effectively swaps $\omega _{1}(r)$ and $\omega _{2}(r)$ from (\ref{eq:su3solgauge}).
In both figures we see a region of solutions in which both gauge field functions have no zeros, but the size of this
region increases for larger $\left| \Lambda \right| $.

\subsubsection{${\mathfrak {su}}(4)$ solitons}
\label{sec:su4solitons}

In ${\mathfrak {su}}(4)$ EYM, the matrix ${\cal {M}}_{N-1}$ (\ref{eq:calM}) reads
\begin{equation}
{\cal {M}}_{3} = \left(
\begin{array}{ccc}
6 & -{\sqrt {12}} & 0 \\
-{\sqrt {12}} & 8 & -{\sqrt {12}} \\
0 & -{\sqrt {12}} & 6
\end{array}
\right) ,
\end{equation}
and has eigenvalues $2$, $6$ and $12$ (\ref{eq:eigenvalues}) and normalized eigenvectors
\begin{eqnarray}
{\bmath {v}}_{1} & = &
\frac {1}{{\sqrt {10}}} \left(
\begin{array}{c}
{\sqrt {3}} \\ 2 \\ {\sqrt {3}}
\end{array}
\right) ; \qquad
{\bmath {v}}_{2} =
\frac {1}{{\sqrt {2}}} \left(
\begin{array}{c}
1 \\ 0 \\ -1
\end{array}
\right) ;
\nonumber \\
{\bmath {v}}_{3} & = &
\frac {1}{\sqrt {5}}  \left(
\begin{array}{c}
1 \\ -{\sqrt {3}} \\ 1
\end{array}
\right) .
\end{eqnarray}
The gauge field functions therefore take the form (\ref{eq:omegabeta}):
\begin{equation}
\left( \begin{array}{c}
\omega _{1} \\ \omega _{2} \\ \omega _{3}
\end{array}
\right)
= \left( \begin{array}{c}
{\sqrt {3}} \\ 2 \\ {\sqrt {3}}
\end{array} \right)
+ \frac {1}{{\sqrt {10}}} \left(
\begin{array}{c}
{\sqrt {3}}\beta _{1} + {\sqrt {5}} \beta _{2} + {\sqrt {2}} \beta _{3} \\
2\beta _{1} - {\sqrt {6}} \beta _{3} \\
{\sqrt {3}} \beta _{1} - {\sqrt {5}} \beta _{2} + {\sqrt {2}} \beta _{3}
\end{array}
\right)  .
\label{eq:su4omegabeta}
\end{equation}
The expressions for the Yang-Mills equations (\ref{eq:YMe}) and $p_{\theta }$ (\ref{eq:ptheta})
are now quite lengthy and so we do not reproduce them here.
We now have a four-dimensional parameter space: $\Lambda $, $b_{1}$, $b_{2}$ and $b_{3}$.

\begin{figure}
\begin{center}
\includegraphics[width=5.7cm,angle=270]{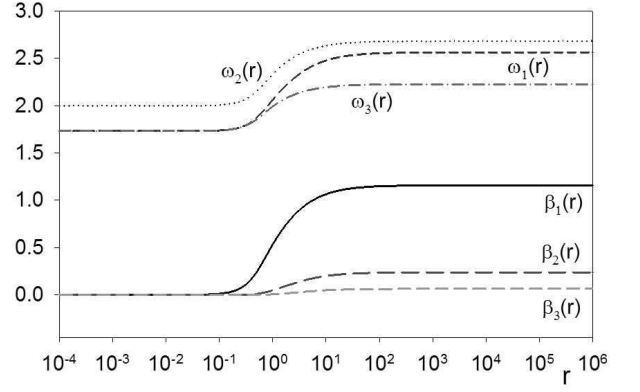}
\end{center}
\caption{Typical ${\mathfrak {su}}(4)$ soliton solution, with $\Lambda = -10$, $b_{1}=-1.2$,
$b_{2} = -0.1$ and $b_{3} = -0.01$.
We plot just the auxiliary functions $\beta _{k}(r)$ and the gauge field functions $\omega _{j}(r)$.
In this example, all three gauge field functions have no zeros.
Note that $\beta _{3}(r)$ is not identically zero, it monotonically increases to $6.41\times 10^{-2}$
as $r\rightarrow \infty $.}
\label{fig:su4solitonsex}
\end{figure}

A typical ${\mathfrak {su}}(4)$ soliton solution is shown in figure \ref{fig:su4solitonsex}.
In figure \ref{fig:su4solitonsex} we have plotted just the gauge field functions $\omega _{j}(r)$ and
the auxiliary functions $\beta _{k}(r)$, as the metric functions have similar behaviour to that seen in,
for example, figure \ref{fig:su3solitonsex}.
In figure \ref{fig:su4solitonsex}, all three gauge field functions have no zeros.
For this large value of $\left| \Lambda \right| $, we find soliton solutions for a wide range of values of
$b_{1}$, all with the three gauge field functions having no zeros.
However, we also find that $b_{2}$ and $b_{3}$ have rather smaller ranges over which we find solutions.
This can be seen in the next two figures.
\begin{figure}
\begin{center}
\includegraphics[width=6.2cm,angle=270]{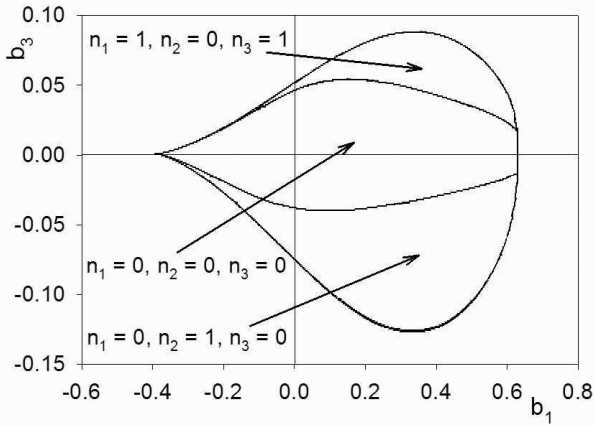}
\end{center}
\caption{Solution space for ${\mathfrak {su}}(4)$ solitons with $\Lambda = -1$ and $b_{2}=0$.
Where we find solutions, the numbers of zeros of the gauge field functions are indicated for the relevant
regions.
For other values of the parameters $b_{1}$ and $b_{3}$, we do not find solutions.
Note that the solution space is {\em {not}} symmetric about the axis $b_{3}=0$.}
\label{fig:su4solitons1}
\end{figure}
\begin{figure}
\begin{center}
\includegraphics[width=6.5cm,angle=270]{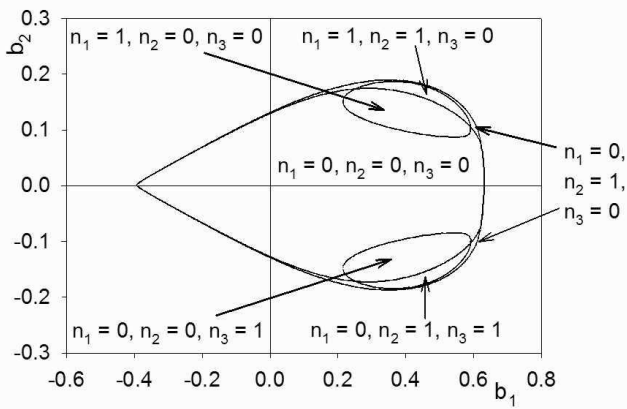}
\end{center}
\caption{Solution space for ${\mathfrak {su}}(4)$ solitons with $\Lambda = -1$ and $b_{3}=0$.
Where we find solutions, the numbers of zeros of the gauge field functions are indicated for the relevant
regions.
For other values of the parameters $b_{1}$ and $b_{2}$, we do not find solutions.
In contrast to figure \ref{fig:su4solitons1}, in this case the solution space is symmetric about
the axis $b_{2}=0$.}
\label{fig:su4solitons2}
\end{figure}

In figures \ref{fig:su4solitons1} and \ref{fig:su4solitons2}, we show the solution spaces of solitons for
$\Lambda = -1$, for $b_{2}=0$ and $b_{3}=0$ respectively.
In each case we have indicated the numbers of zeros of the gauge field functions in those regions where
we find solutions.
Elsewhere, no solutions are found.
The solution space in figure \ref{fig:su4solitons1} is {\em {not}} symmetric about the axis $b_{3}=0$,
but, in figure \ref{fig:su4solitons2} the solution space {\em {is}} symmetric about the axis $b_{2}=0$.
This is expected from the form of the gauge field functions $\omega _{j}(r)$ in terms of the
auxiliary functions $\beta _{k}(r)$ (\ref{eq:su4omegabeta}).
It will be seen from figures \ref{fig:su4solitons1} and \ref{fig:su4solitons2} that, for this value of $\Lambda $,
we have many solutions in which all three gauge field functions have no zeros.

\section{Conclusions}
\label{sec:conc}

In this paper we have presented new soliton and hairy black hole solutions of ${\mathfrak {su}}(N)$ Einstein-Yang-Mills
theory with a negative cosmological constant.
Our solutions are purely magnetic, so that the gauge field functions are in general described by $N-1$ functions,
giving $N-1$ independent degrees of freedom.
This gives, in total, $N+1$ parameters ($N-1$ from the gauge fields, plus the cosmological constant $\Lambda $
and the event horizon radius $r_{h}$, the latter being zero for soliton solutions) which characterize the solutions.
We have developed the formalism for finding solutions for arbitrary $N$, and have discussed in some detail the
properties of the solution space for $N=3$, $4$.

Although the spaces of solutions get progressively more complicated as $N$ increases, the key features are those
also found in the ${\mathfrak {su}}(2)$ case, namely
\begin{enumerate}
\item
Solutions exist in continuous open subsets of the phase space;
\item
As $\left| \Lambda \right| \rightarrow 0$, the solution space fragments and approaches the discrete
solution space in asymptotically flat space;
\item
For sufficiently large $\left| \Lambda \right| $, we have solutions in which all $N-1$ gauge field functions
have no zeros.
\end{enumerate}
The last item is of particular interest.
The existence of these solutions, for sufficiently large $\left| \Lambda \right| $, can be proved analytically \cite{BW}.
In the ${\mathfrak {su}}(2)$ case, it is known that at least some of the solutions in which the gauge field function
has no zeros, for sufficiently large $\left| \Lambda \right| $, are linearly stable, both under spherically symmetric
\cite{ew1,Bjoraker} and non-spherically symmetric \cite{Sarbach} linear perturbations.
We have seen how ${\mathfrak {su}}(2)$ solutions can be embedded into ${\mathfrak {su}}(N)$ EYM, and the first question
is whether those solutions which are stable as solutions of ${\mathfrak {su}}(2)$ EYM remain stable when considered
as solutions of ${\mathfrak {su}}(N)$ EYM.
We will show in a separate publication that this is indeed the case \cite{BW}, and, furthermore, that there
are genuinely ${\mathfrak {su}}(N)$ solutions, in a neighborhood of these embedded ${\mathfrak {su}}(2)$
solutions, which are also stable under linear, spherically symmetric perturbations.
The analysis is rather involved so we do not describe it further here.
The question of non-spherically symmetric perturbations remains open at this stage.

Other interesting open questions remain.
Firstly, there is evidence \cite{suinf} that solutions to ${\mathfrak {su}}(\infty )$ exist in adS, at least for
sufficiently large $\left| \Lambda \right| $.
The field equations for ${\mathfrak {su}}(\infty )$ are rather different in structure from those for
${\mathfrak {su}}(N)$, with the infinite number of ordinary differential YM equations being replaced by
a partial differential equation \cite{suinf}.
Therefore different numerical techniques will be required to solve the field equations.
However, the fact (to be proved in \cite{BW}) that there are soliton and hairy black hole solutions
in ${\mathfrak {su}}(N)$ EYM for any $N$ suggests that non-trivial solutions of ${\mathfrak {su}}(\infty )$ EYM
may indeed exist.
This leaves open the interesting possibility of giving a black hole infinite amounts of gauge field hair.
Secondly, we have not examined the question of whether there are topological black hole solutions of
${\mathfrak {su}}(N)$ EYM, generalizing the topological ${\mathfrak {su}}(2)$ black holes found in \cite{top},
but we anticipate that such solutions exist.
All $k=0$ ${\mathfrak {su}}(2)$ EYM topological black holes are known to be stable as are at least
some of the $k=-1$ solutions \cite{top}, so the stability of any ${\mathfrak {su}}(N)$ EYM topological
black holes would also be of particular interest.
Finally, there is the question of the implication of our solutions for the adS/CFT correspondence \cite{Maldacena}.
A black hole with a particular mass and magnetic charge measured at infinity in adS can now be either an
abelian, magnetically-charged, Reissner-Nordstrom-adS black hole or any one of a number of ${\mathfrak {su}}(N)$
EYM black holes with different $N$.
We would expect that, in analogy with the ${\mathfrak {su}}(2)$ case \cite{SUGRA}, there are ${\mathfrak {su}}(N)$
solutions in some super-gravity theories, which will be even more puzzling in the context of adS/CFT.
It would also be interesting to study the corresponding picture in higher dimensions \cite{higherdim}.
We hope to return to these issues in the near future.

\bigskip
\bigskip

\begin{acknowledgments}
We are grateful to Eugen Radu for innumerable enlightening discussions.
The work of JEB is supported by UK EPSRC, and
the work of EW is supported by UK PPARC, grant reference number PPA/G/S/2003/00082.
\end{acknowledgments}

\end{document}